\documentclass[%
 reprint,
 amsmath,amssymb,
 aps,
 pre,
]{revtex4-2}

\usepackage{graphicx}
\usepackage{dcolumn}
\usepackage{bm}
\usepackage{url}
\usepackage[hidelinks]{hyperref}

\newcommand{\ee}{\mathrm{e}}
\newcommand{\dd}{\; \mathrm{d}}
\newcommand{\ii}{\mathrm{i}}

\begin{document}

\preprint{APS/123-QED}

\title{Flexible String Model of Unsaturated Lipid Bilayer}

\author{Boris Kheyfets}
 \email{kheyfboris@misis.ru}
\author{Sergei Mukhin}

\affiliation{Theoretical Physics and Quantum Technologies Department, National University of Science and Technology MISIS}

\date{\today}

\begin{abstract}
  An analytically solvable model of unsaturated lipid bilayer is derived by introducing finite bending angle of the unsaturated bond relative to straight part of the lipid chain considered previously in our model of semi-flexible strings. It is found that lateral pressure profile of unsaturated lipids has distinct maximum in the unsaturated bond region due to enhanced excluded volume effect caused by the bent bond, leading to an increase of entropic repulsion between the lipid chains. Simultaneously, just away from the unsaturated bond some parts of the neighbouring lipid chains have less probability to collide by geometrical reasons, causing depletion of entropic repulsion relative to saturated lipid chains case and resulting in the local minima of the lateral pressure profile surrounding the maximum at the bent unsaturated bond. Lipid chain order parameter, lateral pressure profile, and area per lipid are computed for POPC and compared with those for DPPC lipid bilayer.
\end{abstract}


\maketitle

\section{
  Introduction
  \label{sec:intro}
}

Consider lipid bilayers that differ only by lipid chain unsaturation: one bilayer consists of saturated lipids, and the other one of unsaturated lipids. Compared to saturated lipid bilayer, unsaturated one would have larger area per lipid (at the same temperature) \cite{kucerka_fluid_2011,patra_under_2006}, lower gel phase transition temperature \cite{kucerka_fluid_2011}, a drop in the chains order parameter \cite{ollila_polyunsaturation_2007,pyrkova_atomic-scale_2011}, and a different lateral pressure profile \cite{ollila_polyunsaturation_2007}. In this paper we reproduce all these characteristic features by making minimal, but pivotal changes to our previously developed analytically solvable flexible string model \cite{mukhin_analytical_2005} of saturated lipid bilayer. The changes are two. The principal one is made in the self-consistent calculation procedure: lipid unsaturation is introduced by allowing only bent chain conformations at the double bond position, while the others are filtered out from the summation over lipid chain conformations  in the partition function of the lipid bilayer. The second change lies in a slight ($\leq 10\%$) tuning of the flexible-string model parameter for incompressible chain's cross-section area of unsaturated lipid relative to saturated one. The latter was done to match equilibrium area per lipid for the unsaturated bilayer found in \cite{kucerka_fluid_2011,patra_under_2006}. 

For reference we used DPPC (16:0) lipids for saturated bilayer, and POPC (18:1/16:0) lipids for unsaturated bilayer. These two cases were chosen as ones of the most well-characterized lipids. POPC is monounsaturated lipid; poly-unsaturated lipids can also be modelled within our model approach, however here we have concentrated solely on DPPC-POPC bilayer differences.

We start with a short overview of the flexible string model for saturated lipid bilayer, including liquid-gel phase transition visible in a free energy plot. These calculations mostly reproduce our previous work \cite{kheyfets_microscopic_2018} with minor simplifications. We then introduce lipid chains unsaturation via specially developed conformational filtering procedure in partition function of the lipids system, which is followed by derivations of analytical expressions for the order parameter and lateral pressure profile. Finally, we apply these approaches to DPPC and POPC models and discuss the results.

\subsection{
  Flexible String Model of Saturated Lipid
  \label{ssec:saturated}
}

\begin{figure}
  \includegraphics[width=0.50\textwidth]{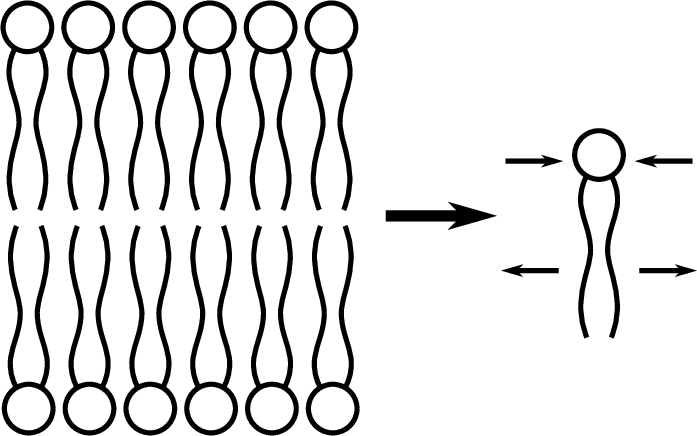}
  \caption{
    Mean-field model of lipid bilayer: bilayer is represented with a single lipid in a mean-field potential. Single chain parameters should be adjusted for the particular bilayer composition and environmental parameters.
    \label{fig:mean-field-model-of-lipid-bilayer}}
\end{figure}

In dynamic mean-field approach, lipid bilayer can be modelled as a single lipid exposed to the self-consistently derived potential, see Fig.~\ref{fig:mean-field-model-of-lipid-bilayer}. Mean-field parameters being derived from the chosen bilayer composition and environment parameters.

\begin{figure}
  \includegraphics[width=0.40\textwidth]{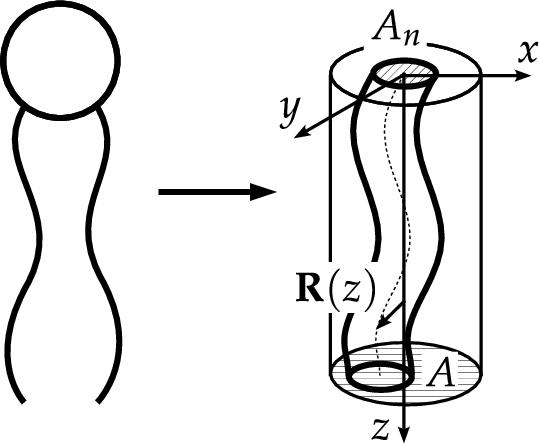}
  \caption{
    Modelling lipid as a single flexible string with bending rigidity $K_f$ and incompressible area $A_n$. On the average a chain swaps the area of $A$.
    \label{fig:flexible-string-model}}
\end{figure}

For a concrete approach we use a model of a single lipid in the mean-field potential that arises from entropic repulsion between fluctuating neighbouring lipid chains due to excluded volume effect \cite{mukhin_analytical_2005}. The minimal model consists of a flexible string in a parabolic mean-field potential, see Fig.~\ref{fig:flexible-string-model}. The energy functional of the string consists of kinetic energy, bending energy, and energy arising from the confining mean-field potential:

\begin{equation}
  \label{eq:ef}
  E_t = \int\limits_0^{L} \left [ \frac{\rho \dot{\mathbf{R}}^2 }{2} + \frac{K_f (\mathbf{R}'')^2}{2} + \frac{B \mathbf{R}^2}{2} \right ] {\, \mathrm{d} z}
  \;,
\end{equation}

\noindent
here ${\bf{R}}(z)=\{R_x(z),R_y(z)\}$ is a vector in the plane of the membrane giving deviation of a string from the $z$ axis (see Fig.~\ref{fig:flexible-string-model}), $\rho$ is a string linear density, $K_f$ is a string bending rigidity, and $B$ is self-consistently derived parameter of entropic repulsion strength, that characterises the parabolic mean-field potential as function of a deviation of the string centres from $z$-axis, $|\mathbf{R}(z)|$, $L$ is a hydrophobic thickness of the monolayer.

Let's find expression for the free energy with the energy functional $E_t$, Eq.~\ref{eq:ef}. For this, we impose boundary conditions on the $\mathbf{R}$, this in turn allows us to represent $E_t$ in the operator form, and finally express free energy $F_t$ in terms of eigenvalues of that operator.

Boundary conditions for a model flexible string take into account the following physical assumptions (same is assumed also for component $R_y(z)$):

  \begin{eqnarray}
   \label{eq:bc}
       &&R_x'(0) = 0 - \nonumber \text{chain director is parallel to} \nonumber\\
       &&\text{ the monolayer normal at membrane surface;}\nonumber
    \\ &&R_x'''(0) = 0 -  \text{net force at chain's head is zero;}
       \\&&R_x''(L) = 0 - \nonumber \text{net torque at chain free end is zero;}
    \\ &&R_x'''(L) = 0 - \nonumber \text{net force at chain free end is zero.}
     \end{eqnarray}

\noindent
The first boundary condition reflects the orientational asymmetry of the monolayer due to the water-lipid interface which is clearly seen from data on the molecules orientational order parameter \cite{lindahl_mesoscopic_2000,vermeer_acyl_2007}: lipid tails are more ordered in the vicinity of head-groups constrained by the hydrophobic tension. Yet, the tilt transition when director forms finite angle with the monolayer normal is not considered here \cite{kheyfets_tilt_2019}. The other boundary conditions reflect the freely moving lipid head-group and hydrocarbon tail end: zero force acting on the head-group and zero torque and force acting on the lipid tail end.

Under the boundary conditions Eq.~\eqref{eq:bc} potential energy part of the functional Eq.~\eqref{eq:ef} can be equivalently rewritten in terms of linear Hermitian operator $\hat H = B + K_f \frac{\partial }{\partial z^4}$ in the form:

\begin{equation}
  \label{eq:of}
  \begin{aligned}
       E_{t(\text{pot})} &= \sum_{\alpha=x,y}E_{\alpha}
    \\ E_{\alpha}      &\equiv \frac{1}{2} \int\limits^{L}_{0}\left[ R_{\alpha}(z)\hat{H}R_{\alpha}(z) \right] {\, \mathrm{d} z}
  \end{aligned}
\end{equation}

\noindent
Then, an arbitrary conformation of the chain can be expressed as the deviation of the centers of the string, $R_{x,y}(z)$, from the straight vertical line (see Fig.~\ref{fig:flexible-string-model}), and is parameterized by a set of coefficients $C_n$ of the linear decomposition of the function $R_{x,y}(z)$ over the eigenfunctions $R_n(z)$ of the operator $\hat{H}$:

\begin{equation}
  \label{eq:eigen1}
  \begin{aligned}
       R_{\alpha=x,y}(z) &= \sum_n C_{n,\alpha}R_n(z)
    \\ \hat{H}R_n(z)  &= E_nR_n(z)
  \end{aligned}
\end{equation}

\noindent
Number of $C_n$ coefficients is related to the number of $CH_2$ groups $N_\mathrm{max}$ in the single chain of the lipid.

Substituting Eq.~\eqref{eq:eigen1} into Eq.~\eqref{eq:ef} and using the standard orthogonality property of the eigenfunctions of operator $\hat{H}$, enables a simple decomposition of the energy functional into the series:
\begin{equation}
  \label{eq:7}
  E_t=\sum_n\frac{1}{2}\left\{ \rho \dot{C}_n^2+E_nC_n^2 \right\}
\end{equation}

\noindent
Hence, Eq.~\eqref{eq:7} reduces the problem of calculating the free energy of the spring to a sum of free energies of non-interecting  1-dimensional harmonic oscillators with "spring rigidities" $E_n$ and  "coordinates" $C_n$.  The corresponding eigenvalues $E_n$ and orthonormalized eigenfunctions $R_n(z)$ of the operator $\hat H$ are \cite{mukhin_analytical_2005}:

\begin{widetext}
\begin{equation}
  \label{eq:eigen2}
  \begin{aligned}
       n &= 0             & \Rightarrow & \left \{
            \begin{aligned}
                 E_0    &= B
              \\ R_0(z) &= \sqrt{\frac{1}{L}}
            \end{aligned}
                                          \right.
    \\ n &\in [1..N_\mathrm{max}] & \Rightarrow & \left \{
            \begin{aligned}
                & E_n   = B + c_n^4 \frac{K_f}{L^4};\;c_n=\pi n-\pi/4;
              \\ &R_n(z) = \sqrt{\frac{2}{L}} \left [ \cos\left (c_n \frac{z}{L} \right ) + \frac{\cos(c_n)}{\cosh(c_n) } \cosh\left (c_n \frac{z}{L} \right ) \right ]
            \end{aligned}
                                                  \right.
  \end{aligned}
\end{equation}
\end{widetext}

\noindent
Arbitrary chain conformation is thus :
$R_x(z) = \sum_{n=0}^{N_\mathrm{max}} R_n(z) C_n$. Due to normalization, $R_n(z)$ dimensionality is $\sim L^{-1/2}$, and hence dimensionality of  $C_n \sim L^{3/2}$.
\noindent
Assuming membrane to be locally isotropic in the lateral plane, we split partition function into a product of two equal components, $Z=Z_xZ_y=Z_x^2$, and thus free energy of the lateral oscillations of the chain equals to \cite{landau_course_statistical}:

\begin{equation}
  \label{eq:Ft}
  F_t = - 2 k_B T \log Z_x; \; Z_x=\prod_n^{N_\mathrm{max}} \frac{k_B T}{\hbar \omega_n};\; \omega_n=\sqrt{\frac{E_n }{\rho}}
  \;.
\end{equation}

\noindent

In order to see how $F_t$ depends on the area per lipid, $A$ (see Fig.~\ref{fig:flexible-string-model}), we note that expression $\partial F_t/\partial  B$ can be computed in two ways - directly via Eqs.(\ref{eq:eigen2},\ref{eq:Ft}):

\begin{equation}
  \label{eq:dFtdB1}
  \frac{\partial F_t}{\partial B} = k_B T \sum_{n=0}^{N_\mathrm{max}} \frac{1}{B + c_n^4 \frac{K_f}{L^4}}
  \;,
\end{equation}

\noindent
and also by representing partition function, $Z_x$ as path integral over all chain conformations, $Z_x = \int \exp(-E_{tx}(R_x) / k_B T)DR_x$:

\begin{eqnarray}
  \label{eq:dFtdB2}
  &&\frac{\partial F_t}{\partial B} = - 2 k_B T \frac{\partial Z_x}{\partial B}{Z_x}^{-1}=\nonumber
  \\ &&=\frac{ \int \left [ \int\limits_0^L R_x^2(z) \dd z \right ] \ee^{\frac{-E_{tx} \lbrace R_x(z) \rbrace}{k_B T}} {\, \mathrm{D} R_x } }{ \int \ee^{\frac{-E_{tx} \lbrace R_x(z)\rbrace}{k_B T}} {\, \mathrm{D} R_x } } =\nonumber\\
  &&= \left \langle \int\limits_0^L R_x^2(z) \dd z  \right  \rangle
  = L \langle R^2 \rangle
  \;,
\end{eqnarray}

\noindent
here $\mathrm{D}$ denotes path integration. Hence,  Eq.\eqref{eq:dFtdB2} shows that $\partial F_t/\partial B$ is proportional to the mean area swept by the centers of the chain. The latter area is also related to the mean area swept by the chain, $A$, and incompressible area of the chain, $A_n$, as: $(\sqrt{A/\pi} - \sqrt{A_n/\pi})^2$. Hence we conclude that:

\begin{equation}
  \label{eq:dFtdB2-1}
  \frac{\partial F_t}{\partial B} = L \frac{\left ( \sqrt{A} - \sqrt{A_n} \right )^2}{\pi}
\end{equation}

Comparing Eq.\eqref{eq:dFtdB1} with Eq.\eqref{eq:dFtdB2-1} one finds:

\begin{equation}
  \label{eq:sc}
  k_B T \sum_{n=0}^{N_\mathrm{max}} \frac{1}{B + c_n^4 \frac{K_f}{L^4}} = L \frac{\left ( \sqrt{A} - \sqrt{A_n} \right )^2}{\pi}
\end{equation}

\noindent
We conclude that $B$ is a monotonically decreasing function of $A$, and hence $F_t$ is also monotonically decreasing function of $A$ (see Eqs.~(\ref{eq:eigen2},\ref{eq:Ft})). Eq.\eqref{eq:sc} constitutes self-consistency equation as it relates mean-field parameter, $B$, with the area per lipid, $A$.

Another contribution to the string's free energy is surface energy, $F_s = \gamma A$, where $\gamma$ is a hydrophobic tension of the lipid membrane, and $A$ is mean area swept by the string. Thus, $F_s$ is monotonically increasing function of mean area swept by the chain. $F_t + F_s$ has a single minimum as a function of $A$, while two minima are needed in order to reproduce liquid to gel phase transition. Hence, in order to get liquid-gel first-order phase transition we need to add one more contribution to the free energy. That contribution is van der Waals interaction between the neighbouring lipids.

Analytical expression for attractive van der Waals interaction between two hydrocarbon chains has been calculated in \cite{salem_attractive_1962} on the basis of quantum mechanics:

\begin{equation}
  \label{eq:vdw}
   F_{VdW} = -U \frac{3\pi}{8L} \frac{N_{CH_2}^2}{D^5}
\end{equation}

\noindent
here $U$ is an interaction constant, and $D$ is a lipid's cross-section diameter.

In the equilibritum free energy of the chain, $F_T = F_t + F_s + F_{VdW}$, is minimal, hence we require that

\begin{equation}
  \label{eq:ec}
  \begin{aligned}
      & \frac{\partial F_T}{\partial A} = 0
   \\ & \Rightarrow \quad
   k_B T \frac{\partial B}{\partial A}\sum_{n=0}^{N_\mathrm{max}} \frac{1}{E_n} + \gamma + \frac{5}{2} \frac{{\tilde U} N_{CH_2}^2}{L A^{7/2}} = 0
  \end{aligned}
\end{equation}

\noindent
Here we used renormalized van der Waals coefficient $\tilde U = 9 \pi^{7/2} U / 2^8$ which arises from using area per lipid instead of lipid diameter, and additional factor of 3 accounts for the fact that every lipid is surrounded by other lipids, assuming a tightly packed lattice, which has a coordination number 6.

\begin{figure}
  \includegraphics[width=0.50\textwidth]{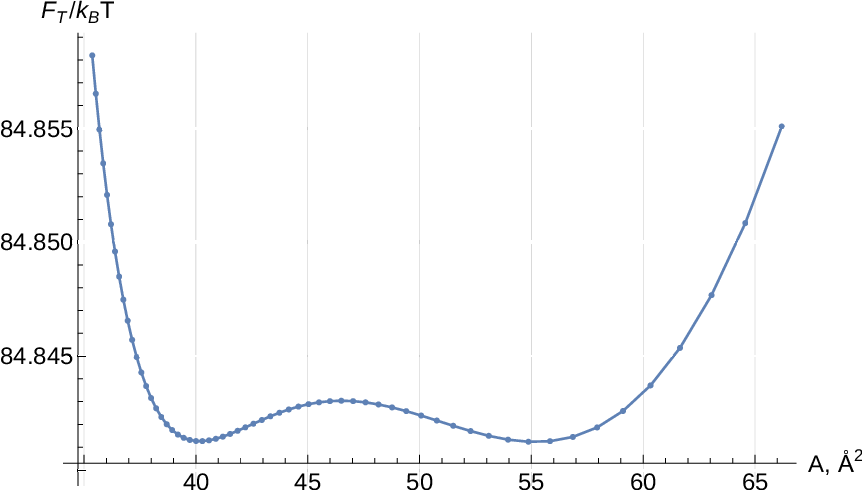}
  \caption{
    Liquid-gel phase transition of DPPC at 314 K. Free energy as function of area per lipid has two minima at the same energy.
    \label{fig:saturated314FT}}
\end{figure}

\begin{figure}
  \includegraphics[width=0.50\textwidth]{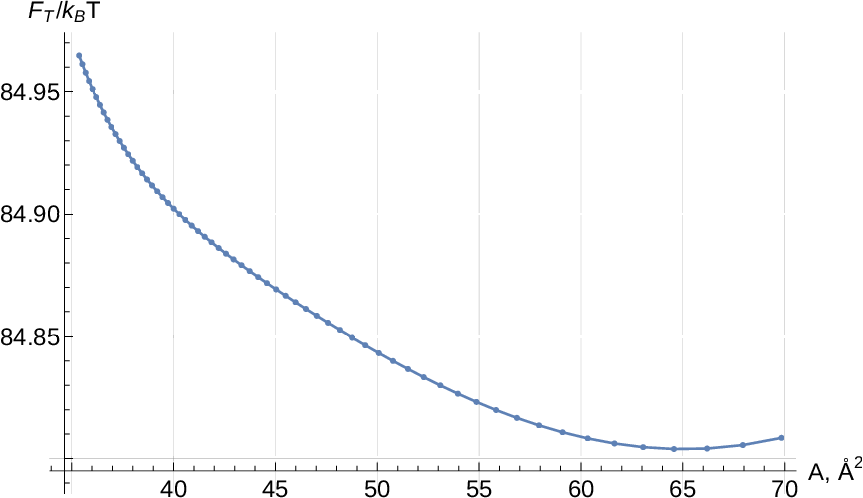}
  \caption{
    Free energy of DPPC as function of area per lipid at 323 K, other parameters are the same as for Fig.~\ref{fig:saturated314FT}.
    \label{fig:saturated323FT}}
\end{figure}

Solving minimum energy condition, Eq.\eqref{eq:ec}, together with self-consistency equation, Eq.\eqref{eq:sc}, for $A$ and $B$ allows to find equilibrium area per lipid at the free energy minima, see Figs.~\ref{fig:saturated314FT} and~\ref{fig:saturated323FT}. We used the following input parameters: incompressible area $A_n = 20$ \AA$^2$, $L = 15$ \AA, $\gamma = 19.37$ erg/cm$^2$, string bending rigidity $K_f = k_B T L/2$, $U = 761$ kcal \AA$^6$/mol, $N_{CH_2} = 16$ number of $CH_2$ groups per chain, $N_\mathrm{max} = 10$ as a maximal number of oscillation modes of the string.

\section{
  Calculation
  \label{sec:calculation}
}

\subsection{
  Flexible String Model of Unsaturated Lipid
  \label{ssec:unsaturated}
}

Let us introduce unsaturated bond in the lipid chain by allowing only bent chain conformations in the partition function of the string. Dropping conformation independent kinetic energy contributions, one finds: 

\begin{eqnarray}
  \label{eq:ZxU}
 &&  Z_x \equiv Z^1_x+Z^2_x=\nonumber\\
 && \frac{1}{2} \int\limits_{-\infty }^{+\infty } \prod_n^{N_\mathrm{max}} \ee^{ - \frac{C_n^2 E_n}{2 k_B T}}
   \cdot \delta \left (
     R_x'(z_0) - C
   \right ){\mathrm{d} C_n}+\nonumber\\
   &&\frac{1}{2} \int\limits_{-\infty }^{+\infty } \prod_n^{N_\mathrm{max}} \ee^{- \frac{C_n^2 E_n}{2 k_B T}}
   \cdot \delta \left (R_x'(z_0) + C\right ) {\mathrm{d} C_n}
\end{eqnarray}

\noindent
Here Dirac delta function $\delta\left (R_x'(z_0) \pm C\right )$ filters out string conformations that do not have a particular bending angle at the depth coordinate $z_0$ inside the bilayer , i.e. at the position of the double bond. $R_x'(z_0) = \tan(\alpha[z_0]) = \sum_{n=0}^{N_\mathrm{max}} C_n R_n'(z_0) = \pm C$, and $\alpha[z_0]$ is tangent angle of the chain at the double bond, $1/2$ takes into account that the chain is bent in either of two sides, i.e.  $\tan(\alpha[z_0])=\pm C$ correspondently,  with the same probability.

For self-consistency equation, Eq.\eqref{eq:dFtdB2-1}, we can concentrate on the parts of the $Z_x$ that include $B$. 

\begin{eqnarray}
  \label{eq:ZxU1}
&& Z^1_x=  \frac{1}{2} \int\limits_{-\infty }^{+\infty } \ee^{- \sum_{n=0}^{N_\mathrm{max}} \frac{C_n^2 E_n}{2 k_B T}}
   \cdot \delta \left (
     \sum_{n=0}^{N_\mathrm{max}} C_n R_n'(z_0) - C
   \right )\times \nonumber\\
   &&\prod_{n=0}^{N_\mathrm{max}} \frac{\mathrm{d}{C}_n}{L^{3/2}}
  \end{eqnarray}

\noindent
where $L^{3/2}$ is introduced to keep the corresponding $Z_x$ part dimensionless. One has to use the following trick to calculate the above path-integral. Namely, the Dirac delta function is expressed using the well known identity:
\begin{equation}
\delta(x)=\int\limits_{-\infty }^{+\infty }\ee^{ixk}\frac{dk}{2\pi}
\label{eq:del}
\end{equation}
\noindent
Hence, substituting the Dirac delta function in Eq.\eqref{eq:ZxU1} by identity from Eq.\eqref{eq:del} and integrating first over all the $ \prod_{n=0}^{N_\mathrm{max}} {\mathrm{d}{C}_n}$ and then over $dk$, one finds:
\begin{eqnarray}
  \label{eq:ZxU2}
&&Z^1_x =\frac{1}{2} \left [  \prod_n \frac{1}{\sqrt{E_n L^3}}\right ]
\cdot \exp\left\{ - \frac{C^2}{2 k_B T\sum\limits_n \frac{\left [R_n'(z_0)\right ]^2}{E_n}}\right\}\times\nonumber\\
&& \frac{1}{ \sqrt{\sum\limits_n \frac{\left [R_n'(z_0)\right ]^2}{E_n}}}=Z^2_x;\; Z_x=2Z^1_x
\end{eqnarray}
\noindent 
where the last equality in Eq. \eqref {eq:ZxU2} follows from the fact that $Z_x^1$ depends only on $C^2$, and hence $Z^1_x=Z^2_x=1/2 Z_x$.
Substituting this into Eq.\eqref{eq:Ft} one finds:

\begin{eqnarray}
  \label{eq:FtU}
&& F_t = k_B T \ln \left ( \left [ \prod\limits_n  E_n L^3 \right ] \sum\limits_n \frac{\left [R_n'(z_0)\right ]^2}{ E_n}\right )+\nonumber\\
  && +\frac{C^2}{\sum\limits_n \frac{\left [R_n'(z_0)\right ]^2}{E_n}}
\end{eqnarray}

Self-consistency equation for unsaturated lipids then reads:

\begin{eqnarray}
  \label{eq:scU}
&&k_B T \left ( \sum\limits_n \frac{1}{B + c_n^4 \frac{K_f}{L^4}} \right )=L \frac{(\sqrt{A} - \sqrt{A_n})^2}{\pi}+\nonumber\\
&&+ \frac{\sum\limits_n \frac{[R_n'(z_0)]^2}{E_n^2}}{\sum\limits_n \frac{[R_n'(z_0)]^2}{E_n}}
\left [k_B T  -  \frac{C^2}{\sum\limits_n \frac{[R_n'(z_0)]^2}{E_n}}\right ]\nonumber\\
\end{eqnarray}

\noindent
(compare with Eq.\eqref{eq:sc}). The equilibrium condition is:

\begin{widetext}
\begin{equation}
  \label{eq:ecU}
  \frac{\partial F_T}{\partial A} = 0
  \quad \Rightarrow \quad
  k_B T \frac{\partial B}{\partial A} \left \lbrace \left ( \sum\limits_n \frac{1}{E_n} \right )
- \frac{ \sum\limits_n \frac{[R_n'(z_0)]^2}{E_n^2}}{\sum\limits_n \frac{[R_n'(z_0)]^2}{E_n}}\left [1-  \frac{C^2 K_f}{k_B T L \sum\limits_n \frac{[R_n'(z_0)]^2}{E_n}}
\right ] \right \rbrace + \gamma + \frac{5}{2} \frac{{\tilde U} N^2}{L A^{7/2}} = 0
\end{equation}
\end{widetext}

\noindent
(compare with Eq.\eqref{eq:ec}).

\begin{figure}
  \includegraphics[width=0.50\textwidth]{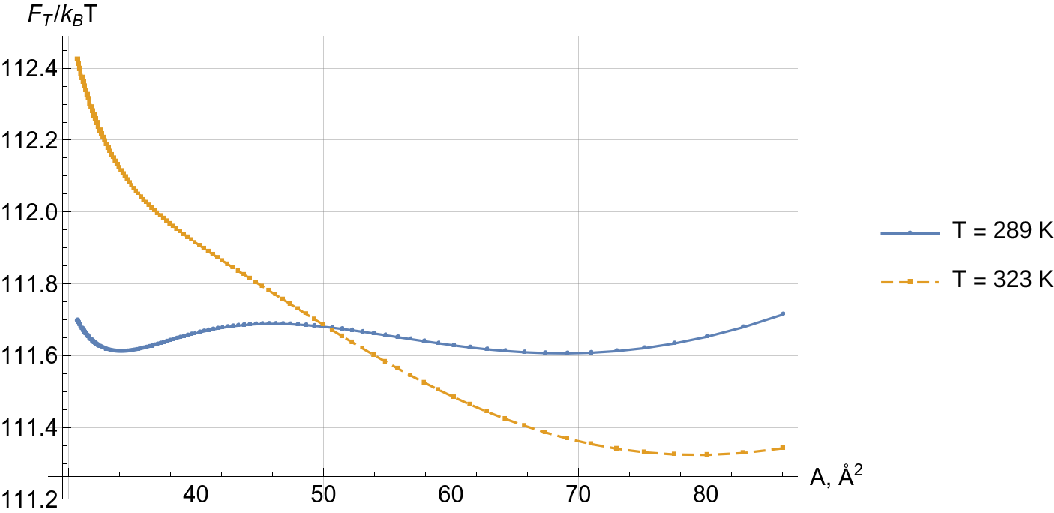}
  \caption{
    Calculated free energy dependences on area per lipid for unsaturated lipid membrane at 289 and 323 K. All parameters, except incompressible string area: $A_n =18$ \AA$^2$ instead of  saturated lipid case $A_n = 20$ \AA$^2$, see text, are the same as in Figs.~\ref{fig:saturated314FT}, \ref{fig:saturated323FT}. In comparison with saturated lipids, unsaturated ones have lower transition temperature into gel phase than in Fig.~\ref{fig:saturated314FT} and larger cross-section area at 323 K than in Fig.~\ref{fig:saturated323FT}.  
  \label{fig:unsaturatedFT}}
\end{figure}

Solving self-consistency condition, Eq.\eqref{eq:scU}, together with equilibrium condition, Eq.\eqref{eq:ecU}, for $A$ and $B$ allows to find equilibrium area per lipid, see Fig.\ref{fig:unsaturatedFT}. All the distinct differences with the saturated lipids case in Figs.~\ref{fig:saturated314FT} and \ref{fig:saturated323FT} are solely due to inclusion of Dirac delta function filter into saturated lipids partition function, Eq.\eqref{eq:ZxU}, where we have used $C = 0.5$ corresponding to double bond bending angle $\alpha = 26.4 \deg$. Slight decrease of incompressible string area for unsaturated lipids: $A_n =18$ \AA$^2$ instead of  saturated lipid case: $A_n = 20$ \AA$^2$ was necessary to keep equilibrium area per lipid in the range found earlier \cite{kucerka_fluid_2011,patra_under_2006}.

Computed equilibrium areas per lipid for POPC shown on Fig.~\ref{fig:unsaturatedFT} are slightly above the values reported in the literature \cite{kucerka_fluid_2011,pabst_x-ray_2000,pabst_structural_2000,patra_under_2006}. This means that one should change some other parameters of the calculation for the closer match with experimental data on area per lipid. However, in this study we focus on a single most important feature of the unsaturated lipids, namely, the double bond, which is introduced by imposing a filter in the partition function, Eq.\eqref{eq:ZxU}. In order to see the effect of this particular change, we had chosen to make as little changes as possible to the other calculation parameters.

\subsection{
  Lateral Pressure Profile
  \label{ssec:lpp}
}

Lateral pressure profile is a function $\Pi(z)$ which is equal to hydrophobic tension, $\gamma$, when integrated over the thickness of the monolayer:

\begin{equation}
  \label{eq:lpp1}
  \int_0^L \Pi(z) {\, \mathrm{d} z} = \gamma
\end{equation}

\noindent
Comparing this with equilibrium condition, Eq.\eqref{eq:ec} or Eq.\eqref{eq:ecU}, we see that it has two components -- one is due to entropic repulsion, $F_t$, and the other is due to van der Waals attraction, $F_{VdW}$:

\begin{equation}
  \label{eq:lpp2}
  \int_0^L \Pi(z) {\, \mathrm{d} z} = {- \frac{\partial F_t}{\partial A}} \quad {- \frac{5}{2} \frac{{\tilde U} N_{CH_2}^2}{L A^{7/2}}}
\end{equation}

For saturated lipids, lateral pressure profile due to entropic repulsion is

\begin{equation}
  \label{eq:lpp3}
  \int_0^L \Pi_t(z) {\, \mathrm{d} z} = - \frac{\partial F_t}{\partial A} = - k_B T \sum_{n=0} \frac{1}{E_n} \frac{\partial E_n}{\partial A}
\end{equation}

\noindent
By definition, $E_n$ and $R_n(z)$ are eigenvalues and eigenfunctions of operator $\hat H$ in Eq. \ref{eq:of}: $\hat H R_n(z) = E_n R_n(z)$. Hence we write

\begin{equation}
  \label{eq:lpp4}
  \begin{aligned}
       & \int_0^L R_n {\hat H} R_n {\, \mathrm{d} z} = \int_0^L R_n E_n R_n {\, \mathrm{d} z} \equiv E_n \int_0^L R_n^2 {\, \mathrm{d} z}
    \\ & \Rightarrow \quad \frac{\partial E_n}{\partial A} = \frac{\partial E_n}{\partial A} \int_0^L R_n^2 {\, \mathrm{d} z}
  \end{aligned}
\end{equation}

\noindent
Comparing this with Eq.\eqref{eq:lpp3} we conclude that

\begin{equation}
  \label{eq:lpp5}
  \begin{aligned}
    \Pi_t(z) &= - k_B T \sum_{n=0} \frac{1}{E_n} {\frac{\partial E_n}{\partial A}}  R_n^2(z)
    \\ &\equiv - k_B T \frac{\partial B}{\partial A} \sum_{n=0} \frac{R_n^2(z)}{E_n} {\frac{\partial E_n}{\partial B}}
    \\ &= - k_B T \frac{\partial B}{\partial A} \sum_{n=0} \frac{R_n^2(z)}{E_n}
  \end{aligned}
\end{equation}

Similarly, for unsaturated lipids, one arrives at the following expression for lateral pressure profile:

\begin{widetext}
\begin{equation}
  \label{eq:lppU}
  \Pi_t(z) = - k_B T \frac{\partial B}{\partial A} \sum_{n=0} R_n^2(z) \left [
    \frac{1}{E_n} - \frac{
      \frac{[R_n'(z_0)]^2}{E_n^2}
    }{
      \sum\limits_m \frac{[R_m'(z_0)]^2}{E_m}
    } \left \lbrace
      1
      -  \frac{
        C^2 K_f
      }{
        k_B T L \sum\limits_m \frac{[R_m'(z_0)]^2}{E_m}
      }
    \right \rbrace
  \right ]
\end{equation}
\end{widetext}

\begin{figure}
  \includegraphics[width=0.50\textwidth]{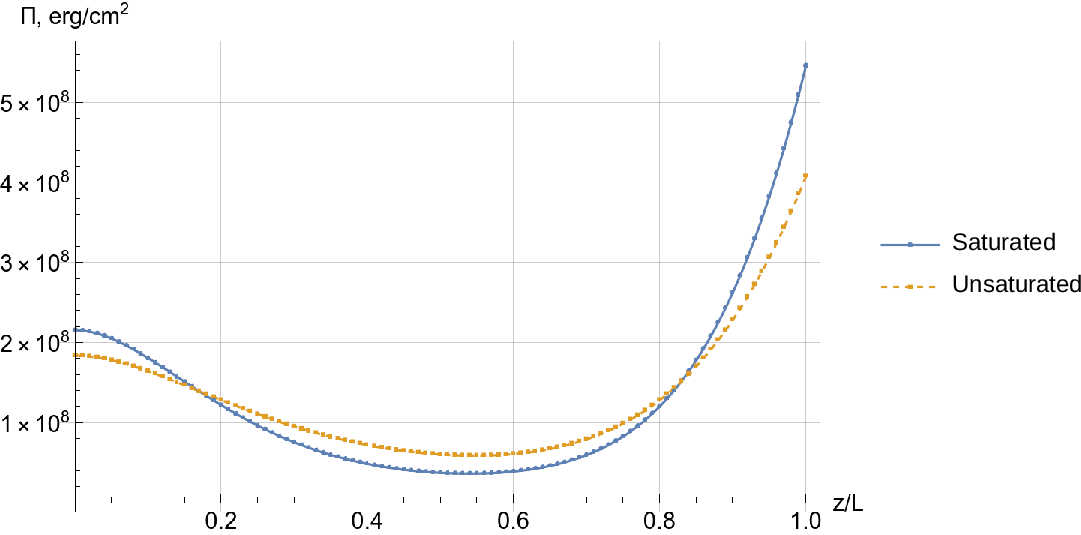}
  \caption{
    Computed lateral pressure profile for DPPC and POPC lipids at 323 K.
    \label{fig:lateral-pressure-profiles}}
\end{figure}

Lateral pressure profile plots for saturated and unsaturated lipids are plotted in Fig.~\ref{fig:lateral-pressure-profiles}. We see that lateral pressure peak at the midplane area ($z/L = 1$) becomes smaller with unsaturation. Lateral pressure profile was studied in \cite{ollila_polyunsaturation_2007} by means of molecular dynamics: in Fig. 7 of  their paper \cite{ollila_polyunsaturation_2007} the authors plot lateral pressure profile of lipids with increasing unsaturation and also find that the peak at the midplane area diminishes with unsaturation.

\subsection{
  Order Parameter
  \label{ssec:op}
}

By definition, molecular order parameter is

\begin{equation}
  \label{eq:op}
  S_\mathrm{mol}(z) = \frac{1}{2} \left (
  3 \langle \cos^2 \theta(z) \rangle - 1
\right )
\end{equation}

In flexible string model we have an approximation for tangent:

\begin{equation}
  \label{eq:tan}
  \tan \theta(z) \approx R_x' = \sum C_n R_n'(z)
\end{equation}

\noindent
hence we express $\cos\theta$ via $\tan\theta$ as:

\begin{equation}
  \label{eq:costan}
  \cos^2 \theta = \frac{1}{1 + \tan^2{\theta}}
\end{equation}

\noindent
Also note that

\begin{equation}
  \label{eq:a}
  \begin{aligned}
    & \frac{1}{1 + a(z)^2} = \frac{1}{2} \int_0^\infty \ee^{-x} \left [  \ee^{\ii a(z) x} + \ee^{-\ii a(z) x}\right ]  {\, \mathrm{d} x}
    \\ & \Rightarrow \left \langle \frac{1}{1 + a(z)^2} \right \rangle
         = \frac{1}{2} \int_0^\infty \ee^{-x} \left [
             \langle \ee^{\ii a(z) x}  \rangle
           + \langle \ee^{-\ii a(z) x} \rangle
         \right ] \mathrm{d} x
  \end{aligned}
\end{equation}

\noindent
Thus, intermediate expression for the order parameter is

\begin{widetext}
  \begin{equation}
    \label{eq:orI}
    S_\mathrm{mol}(z) = \frac{1}{2} \left [
      3 \int_0^\infty \ee^{-x} \left [
        \left \langle
          \ee^{\ii x \sum C_n R_n'(z)}
        \right \rangle
        + \left \langle
          \ee^{-\ii x \sum C_n R_n'(z)}
        \right \rangle
      \right ] {\, \mathrm{d} x}
      - 1
    \right ]
  \end{equation}
\end{widetext}

Starting with saturated lipids, we write:

\begin{equation}
  \label{eq:op1}
  \left \langle
    \ee^{\ii \cdot \sum C_n R_n'(z) \cdot x}
  \right \rangle = \frac{\int \ee^{\ii \cdot \sum C_n R_n'(z) \cdot x} \cdot \ee^{-\frac{\sum C_n^2 E_n}{2 k_B T}}
    \prod {\, \mathrm{d} C_n}}{\int \ee^{-\frac{\sum C_n^2 E_n}{2 k_B T}}\prod {\, \mathrm{d} C_n}}
\end{equation}

The numerator integral here is:

\begin{widetext}
  \begin{equation}
    \label{eq:op2}
       \int \ee^{
        - \sum\limits_n \frac{E_n}{2 k_B T} \left [
        C_n^2 - 2 C_n \ii x R_n' \frac{k_B T}{E_n} + \left (
        \ii x R_n' \frac{k_B T}{E_n}
        \right )^2
        - \left (
        \ii x R_n' \frac{k_B T}{E_n}
        \right )^2
        \right ]
        } \prod {\, \mathrm{d} C_n}
      = \ee^{
           - \frac{x^2 k_B T}{2} \sum\limits_n \frac{(R_n')^2}{E_n}
           } \prod \sqrt{
           \frac{\pi \cdot 2 k_B T}{E_n}
           }
  \end{equation}
\end{widetext}

\noindent
The denominator integral is the same, but with $x=0$. Hence, one obtains:

\begin{equation}
  \label{eq:op3}
  \left \langle
  \ee^{\ii \cdot \sum C_n R_n'(z) \cdot x}
  \right \rangle
  = \ee^{
    - \frac{x^2 k_B T}{2} \sum\limits_n \frac{(R_n')^2}{E_n}
  }
\end{equation}

\noindent
Thus, we see that expression in Eq.\eqref{eq:op1} is real, hence it coincides with its complex conjugate, so we conclude that Eq.\eqref{eq:a} for saturated lipids reads:

\begin{equation}
  \label{eq:op4}
  \begin{aligned}
    \left \langle  \frac{1}{1 + a(z)^2}\right \rangle &= \frac{1}{2} \int_0^\infty \ee^{-x}
    \ee^{-x^2 \frac{k_B T}{2} \sum\limits_n \frac{(R_n')^2}{E_n}}
    {\, \mathrm{d} x}
    \\ &= \frac{\ee^{\frac{1}{4y(z)}} \sqrt{\pi} \, \mathrm{erfc}\left (
    \frac{1}{2 \sqrt{y(z)}}
    \right )}{2 \sqrt{y(z)}}
  \end{aligned}
  \end{equation}
  \noindent
where:
\begin{equation}
 \label{eq:yz}
y(z)=\frac{k_B T}{2} \sum\limits_n \frac{(R_n')^2}{E_n}
 \end{equation}
\noindent
and order parameter for saturated lipids is

\begin{equation}
  \label{eq:op5}
S_\mathrm{mol}(z) = \frac{1}{2} \left (
  3 \frac{\sqrt{\pi} \ee^{\frac{1}{4 y(z)}} \mathrm{erfc}\left ( \frac{1}{2 \sqrt{y(z)}} \right ) }{2 \sqrt{y(z)}}
  - 1
\right )
\end{equation}

Similarly, for unsaturated lipids we start with $\langle \ee^{\ii a(z) x} \rangle$:

\begin{equation}
  \label{eq:opU1}
  \begin{aligned}
& \left \langle
\ee^{\ii \cdot \sum C_n R_n'(z) \cdot x}
\right \rangle =
\\ & \frac{
  \int \ee^{\ii \cdot \sum C_n R_n'(z) \cdot x} \cdot \ee^{-
    \frac{
      \sum C_n^2 E_n
    }{
      2 k_B T
    }
  }
  \cdot \delta\left [
    \sum C_n R_n'(z_0) - C
  \right ]
  \prod {\, \mathrm{d} C_n}
}{
  \int \ee^{-
    \frac{
      \sum C_n^2 E_n
    }{
      2 k_B T
    }
  }
  \cdot \delta\left [
    \sum C_n R_n'(z_0) - C
  \right ]
  \prod {\, \mathrm{d} C_n}
}
  \end{aligned}
\end{equation}

\noindent
Consider the numerator integral first. Using Dirac delta function representation in Eq.\eqref{eq:del}, completing the square for $C_n$, integrating over  $C_n$ and then again completing the square for $k$ and integrating over $k$ afterwards one finds:

\begin{widetext}
\begin{equation}
  \label{eq:opU2}
  \begin{aligned}
  & \int \ee^{\ii \cdot \sum C_n R_n'(z) \cdot x} \cdot \ee^{-
    \frac{
      \sum C_n^2 E_n
    }{
      2 k_B T
    }
  }
  \cdot \delta\left [
    \sum C_n R_n'(z_0) - C
  \right ]
  \prod {\, \mathrm{d} C_n} = \frac{1}{2 \pi}
\left ( \prod \sqrt{\frac{\pi \cdot 2 k_B T}{E_n}} \right )
\cdot \sqrt{
  \frac{2 \pi}{ k_B T \sum \frac{[R_n'(z_0)]^2}{E_n} }
}
  \\ &
\cdot \ee^{
  - \frac{
    C^2
  }{
    2 k_B T \sum \frac{[R_n'(z_0)]^2}{E_n}
  }
}
\cdot \ee^{
  \frac{k_B T}{2}
  \frac{
    x^2 \left [
  \sum \frac{R_n'(z_0) R_n'(z)}{E_n}
    \right ]^2
    + \frac{2 x \ii C}{k_B T}\sum \frac{R_n'(z_0) R_n'(z)}{E_n}
  }{
    \sum \frac{[R_n'(z_0)]^2}{E_n}
  }
}
\cdot \ee^{
  - \frac{k_B T}{2} x^2 \sum \frac{[R_n'(z)]^2}{E_n}
  }
  \end{aligned}
\end{equation}
\end{widetext}

The denominator integral in Eq.\eqref{eq:opU1} will be equal to the one in Eq.\eqref{eq:opU2} with $x=0$. Thus:

\begin{widetext}
\begin{equation}
  \label{eq:opU3}
  \begin{aligned}
\left \langle \ee^{\ii a(z) x} \right \rangle &= \ee^{
  \frac{k_B T}{2}
  \frac{
    x^2 \left [
  \sum \frac{R_n'(z_0) R_n'(z)}{E_n}
    \right ]^2
    + \frac{2 x \ii C}{k_B T}\sum \frac{R_n'(z_0) R_n'(z)}{E_n}
  }{
    \sum \frac{[R_n'(z_0)]^2}{E_n}
  }
}
\cdot \ee^{
  - \frac{k_B T}{2} x^ 2\sum \frac{[R_n'(z)]^2}{E_n}
  }
\\ &= \ee^{
     -x^2 {
\frac{
     k_B T
     }{
     2
     }
     \left \lbrace
     \sum \frac{[R_n'(z)]^2}{E_n}
     - \frac{
      \left (
  \sum \frac{R_n'(z_0) R_n'(z)}{E_n}
    \right )^2
      }{
      \sum \frac{[R_n'(z_0)]^2}{E_n}
      }
\right \rbrace
     }
     +\ii {
     Cx
          \frac{
              \sum \frac{R_n'(z_0) R_n'(z)}{E_n}
     }{
     \sum \frac{[R_n'(z_0)]^2}{E_n}
     }
     }
     }
  \end{aligned}
\end{equation}
\end{widetext}

\noindent
Mean for complex conjugate, $\langle \ee^{-\ii a(z) x} \rangle$, will be the same, but with changing $x$ to $-x$ in Eq.\eqref{eq:opU3}. Sum of these two is real:

\begin{eqnarray}
  \label{eq:opU4}
&&  \ee^{
  -y(z) x^2 + \ii w(z) x
}
+ \ee^{
  -y(x) x^2 - \ii w(z) x
}
= \nonumber\\
&&=2 \ee^{
  -y(z) x^2
}
\cos w(z) x
\end{eqnarray}
\noindent
where now: 
\begin{equation}
\label{eq:yofz}
{y(z)}={
\frac{
     k_B T
     }{
     2
     }
     \left [
     \sum \frac{[R_n'(z)]^2}{E_n}
     - \frac{
      \left (
  \sum \frac{R_n'(z_0) R_n'(z)}{E_n}
    \right )^2
      }{
      \sum \frac{[R_n'(z_0)]^2}{E_n}
      }
\right]
     }
\end{equation}

\begin{equation}
\label{eq:wofz}
{w(z)}={
     C
          \frac{
              \sum \frac{R_n'(z_0) R_n'(z)}{E_n}
     }{
     \sum \frac{[R_n'(z_0)]^2}{E_n}
     }
     }
 \end{equation}
\noindent
Then, Eq.\eqref{eq:a} for unsaturated lipids is:

\begin{equation}
  \label{eq:opU4}
\int_0^\infty \ee^{
  -y(z) x^2 - x
}
\cos w(z) x
{\, \mathrm{d} x}
\end{equation}

\noindent
Calculating this integral we finally arrive at the expression for order parameter of unsaturated lipids:

\begin{widetext}
\begin{equation}
  \label{eq:opU5}
     S_\mathrm{mol}(z) =  \frac{1}{2} \left (
  3 \frac{
    \sqrt{\pi} \ee^{- \frac{(w(z) + \ii)^2}{4 y(z)}} \left [
      1 + \ee^{\frac{\ii w(z)}{y(z)}} \mathrm{erfc}\left ( \frac{1 + \ii w(z)}{2\sqrt{y(z)}}\right ) + \ii \, \mathrm{erfi}\left ( \frac{\ii + w(z)}{2 \sqrt{y(z)}} \right )
    \right ]
  }{
    4 \sqrt{y(z)}
  }
  - 1
\right )
\end{equation}
\end{widetext}

\begin{figure}
  \includegraphics[width=0.50\textwidth]{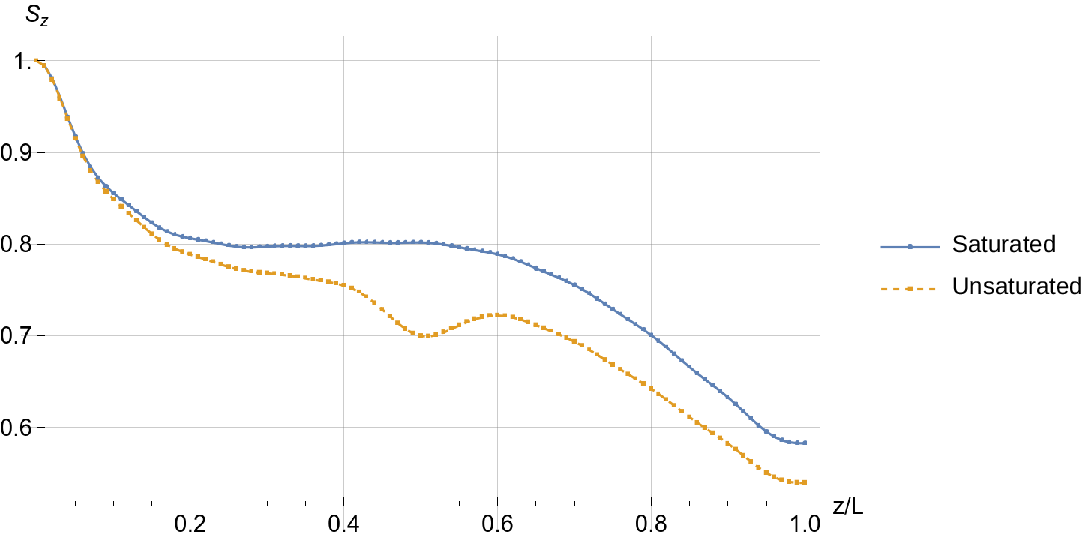}
  \caption{
    Plots of calculated order parameter distribution using Eq.\eqref{eq:op5} for saturated lipids corresponding to DPPC case, and Eq.\eqref{eq:opU5} for unsaturated lipids corresponding to POPC case, at temperature $T=323$ K.
    \label{fig:order-parameters}}
\end{figure}

Order parameter plots for saturated Eq.\eqref{eq:op5} and unsaturated Eq.\eqref{eq:opU5} lipids are plotted in Fig.~\ref{fig:order-parameters}. The order parameter curve Eq.\eqref{eq:opU5}, corresponding to e.g. the case of unsaturated POPC, has a drop in the middle of the monolayer, where double bond is located at $z_0=0.5$ in our model. This result coincides with the molecular dynamics study, \cite{ollila_polyunsaturation_2007} (see Fig. 10 therein) and \cite{pyrkova_atomic-scale_2011} (see Fig. 2 therein).

In passing, we note that order parameter for DPPC and POPC has also been reported in \cite{patra_under_2006} (Fig. 13 therein), but the authors had plotted order parameter only for saturated POPC chain, which closely matches order parameter for DPPC.

\section{
  Discussion
  \label{sec:discussion}
}

To summarise, we have modelled  unsaturated POPC lipid in our flexible strings model by allowing only strings conformations that possess fixed nonzero tangent angle of the double bond with respect to preceding it saturated bonds at some definite position $z_0$ inside the bilayer, e.g. $\tan[\alpha(z_0=0.5 L)] = 0.4, 0.6, ...$ etc. This results in larger area per lipid compared to saturated lipids, as e.g. in DPPC lipid, and produces lateral pressure profiles and order parameter distributions that match qualitatively the ones reported in \cite{ollila_polyunsaturation_2007}. The most important physics conclusion that follows from our results is obvious from Figs.~\ref{fig:lateral-pressure-profiles}, \ref{fig:lateral-pressure-profile-z0}. Namely, the lateral pressure profiles of unsaturated lipids near the unsaturated bonds go beyond saturated lipids profile, as would be expected from considerations of excluded volume effect on the entropic repulsion between the lipid chains. The unsaturated bond has finite bending angle with respect to straight line of the lipid chain and, therefore, increases locally the lipid occupied area, thus causing stronger entropic repulsion between the lipid chains. Simultaneously, away from the unsaturated (bent) bond the neighbouring lipid chains have less probability to collide just by geometrical reasons, and therefore, the entropic repulsion between them is depleted relative to saturated lipid chains case, thus causing relative decrease of the lateral pressure profile with respect to saturated lipids case in these same regions away from the bent unsaturated bond, as is seen in Figs.~\ref{fig:lateral-pressure-profiles}, \ref{fig:lateral-pressure-profile-z0}.

In support of the above reasoning we also have checked predictions of our model for the cases of unsaturated double bond with different values of the bending angle and with different positions of the unsaturated bond along the lipid chains.

\begin{figure}
  \includegraphics[width=0.50\textwidth]{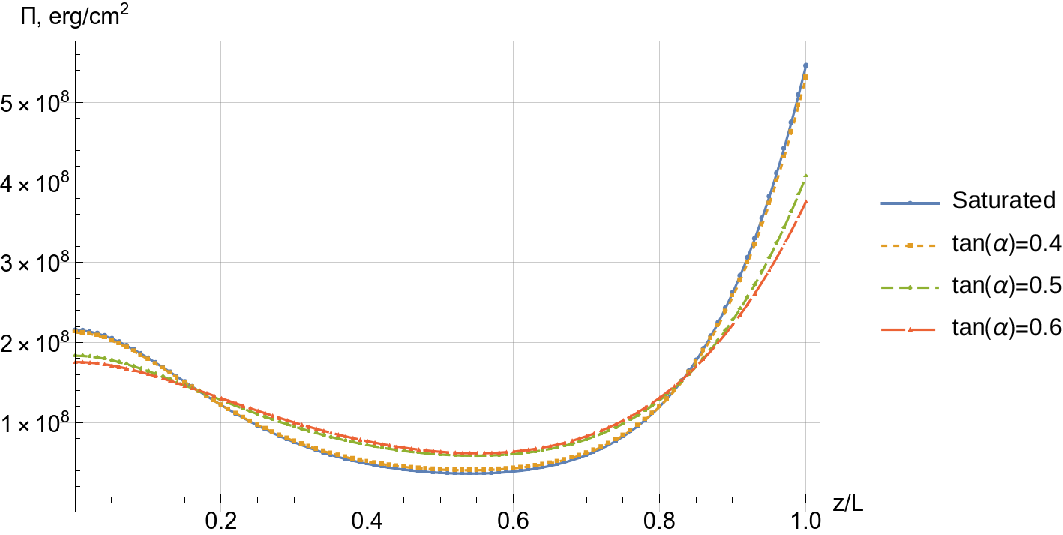}
  \caption{
    Calculated lateral pressure profiles for various values of the fixed double bond tangent angle at 323 K. In all cases the tangent angle is fixed at the position $z_0 = 0.5 L$ inside a monolayer of thickness $L$.
    \label{fig:lateral-pressure-profile-angles}}
\end{figure}

Let's start with different values of the double bond angle. Presented here Fig.~\ref{fig:lateral-pressure-profile-angles} shows lateral pressure profiles for various values of the fixed double bond tangent angle calculated in our model,  see Eq. \eqref{eq:lppU}. We see that with decreasing double bond tangent angle the lateral pressure profile approaches the one of the saturated lipid, calculated using Eq. \eqref{eq:lpp5}.

\begin{figure}
  \includegraphics[width=0.50\textwidth]{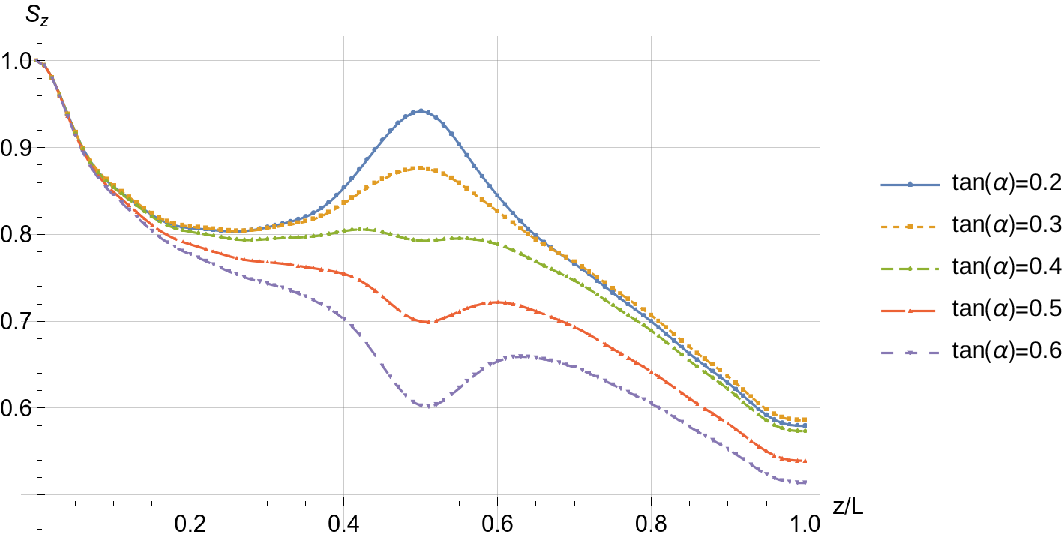}
  \caption{
    Calculated order parameter for various values of the fixed double bond angle at 323 K. In all cases the angle is fixed at $z_0 = 0.5 L$. For $\tan(\alpha)$ values of $0.2$ and $0.3$ we had to increase incompressible area, $A_n$, to $23 \AA^2$ and $20 \AA^2$ respectively to prevent transition into gel-phase. 
    \label{fig:order-parameter-angles}}
\end{figure}

\begin{figure}
  \includegraphics[width=0.50\textwidth]{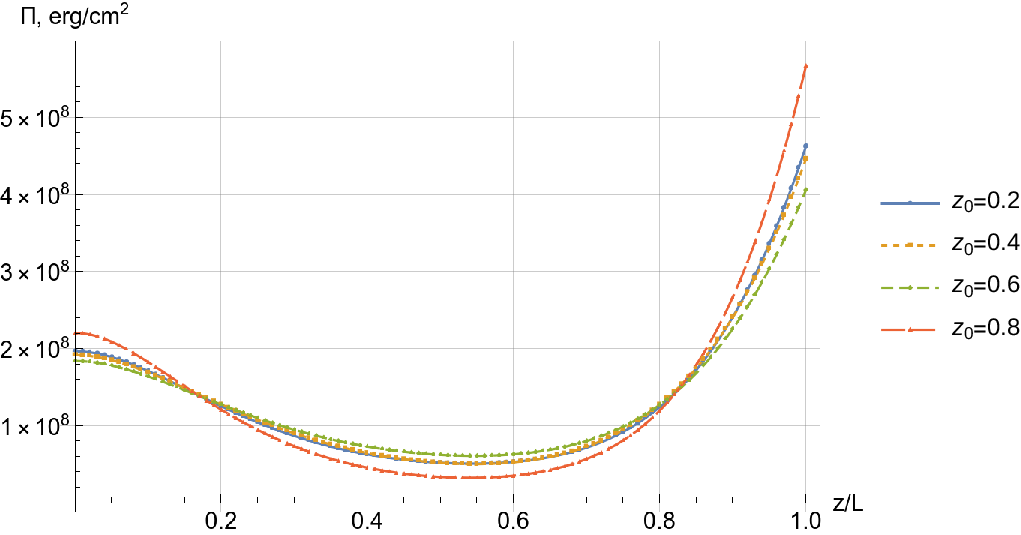}
  \caption{
    Calculated lateral pressure profiles for various double bond positions at 323 K. In all cases the value of the double bond tangent angle is fixed at $\tan(\alpha) = 0.5$. For $z_0$ values of $0.2$ and $0.8$ we had to increase incompressible area, $A_n$, to $20 \AA^2$ and $19 \AA^2$ respectively to prevent transition into gel-phase.
    \label{fig:lateral-pressure-profile-z0}}
\end{figure}

Next, we consider effect of double bond tangent angle on the behaviour of the order parameter for various values of the angle. Results of calculations using Eq. \eqref{eq:opU5}are presented in Fig.~\ref{fig:order-parameter-angles}. We see that for angles less then $\tan(\alpha[z_0=0.5 L]) \approx 0.45$ at $z_0=0.5 L$ the order parameter  exceeds the one of the saturated lipid ($S_\mathrm{mol}(z)\approx 0.8$), see Fig.~\ref{fig:order-parameters}. This means, that average tangent value of an angle of saturated lipid at $z=0.5 L$ is about $0.45$.

Farther, we investigated an influence of the position of the double bond on the membrane characteristics. Lateral pressure profiles calculated for various double bond positions using Eq. \eqref{eq:lppU} are exhibited  in Fig.~\ref{fig:lateral-pressure-profile-z0}. Comparison of the latter with the lateral pressure profile of the saturated lipid in Fig.~\ref{fig:lateral-pressure-profiles} supports proposed above relation between position of unsaturated bond and local excluded volume effect. Also, we see that in our model the closer double bond is to the headgroup of the lipid, the less is its impact on the shape of the lateral pressure profile. This is clearly due to the fixed angle at the headgroup, $R'(0)=0$, see Eq.\eqref{eq:bc}.

\begin{figure}
  \includegraphics[width=0.50\textwidth]{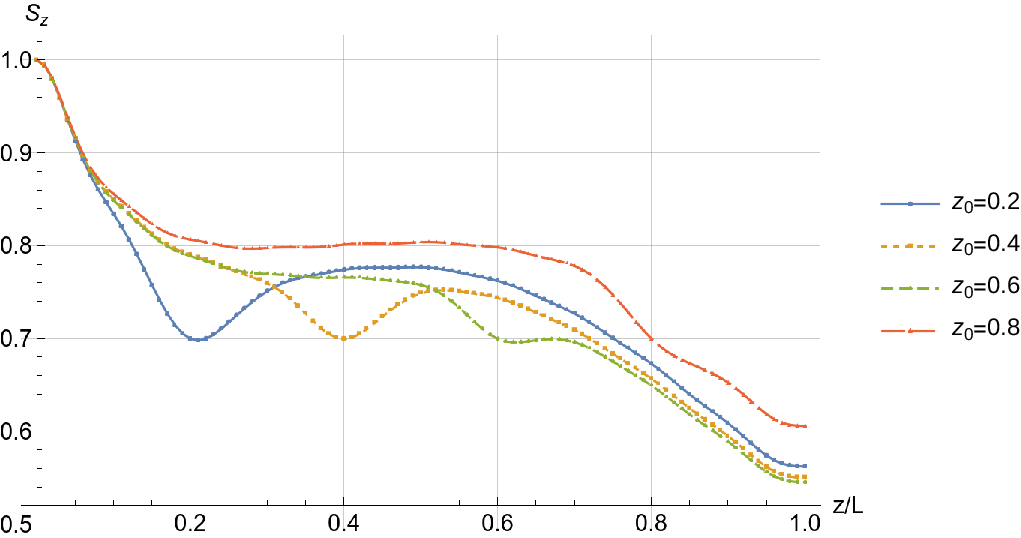}
  \caption{
    Calculated order parameter for various double bond positions $z_0$ at 323 K. In all cases the value of the double bond tangent is fixed at $\tan(\alpha) = 0.5$. For $z_0$ values of $0.2$ and $0.8$ we had to increase incompressible area, $A_n$, to $20 \AA^2$ and $19 \AA^2$ respectively to prevent transition into gel-phase.
    \label{fig:order-parameter-z0}}
\end{figure}

Computed order parameter distributions for various values of the double bond position using Eq.\eqref{eq:opU5} are shown in Fig.~\ref{fig:order-parameter-z0}. Comparing the latter with the order parameter distribution of the saturated lipid, Fig.~\ref{fig:order-parameters}, we see that the double bond position perturbs order parameter locally, so that before and after the double bond the order parameter curve gradually converges to the one of the saturated lipid.

\section{
Acknowledgements 
  \label{sec:acknowldg}
}
Authors acknowledge support by the Federal Academic Leadership Program Priority 2030 (NUST MISIS Grant No.
K2-2022-025).

\bibliography{paper}

\end{document}